\documentclass[letterpaper]{article} % DO NOT CHANGE THIS
\usepackage{aaai25}  % DO NOT CHANGE THIS[submission]
\usepackage{times}  % DO NOT CHANGE THIS
\usepackage{helvet}  % DO NOT CHANGE THIS
\usepackage{courier}  % DO NOT CHANGE THIS
\usepackage[hyphens]{url}  % DO NOT CHANGE THIS
\usepackage{graphicx} % DO NOT CHANGE THIS
\usepackage{natbib}  % DO NOT CHANGE THIS AND DO NOT ADD ANY OPTIONS TO IT
\usepackage{caption} % DO NOT CHANGE THIS AND DO NOT ADD ANY OPTIONS TO IT
\nocopyright
\usepackage{algorithm}
\usepackage{algorithmic}

\usepackage{amsmath,amsfonts,bm,mathtools}
\usepackage{amssymb}
\usepackage{amsthm}
\usepackage{booktabs}
\usepackage{enumitem}
\usepackage{wasysym}
\usepackage{tabularx}

\usepackage[capitalize,noabbrev]{cleveref}

\usepackage{amsmath,amsfonts,bm,mathtools}

\def\eqref#1{equation~\ref{#1}}

\def\1{\bm{1}}

\def\rf{{\textnormal{f}}}

\def\rv{{\textnormal{v}}}

\def\rvu{{\mathbf{i}}}

\def\rvu{{\mathbf{u}}}

\def\rvx{{\mathbf{x}}}

\def\rmI{{\mathbf{I}}}

\def\rmV{{\mathbf{V}}}

\def\vzero{{\bm{0}}}
\def\vone{{\bm{1}}}
\def\vmu{{\bm{\mu}}}

\def\vepsilon{{\bm{\epsilon}}}

\def\vd{{\bm{d}}}
\def\ve{{\bm{e}}}

\def\vr{{\bm{r}}}

\def\vu{{\bm{u}}}
\def\vv{{\bm{v}}}

\def\vx{{\bm{x}}}

\def\vz{{\bm{z}}}

\def\mLambda{{\bm{\Lambda}}}

\DeclareMathAlphabet{\mathsfit}{\encodingdefault}{\sfdefault}{m}{sl}
\SetMathAlphabet{\mathsfit}{bold}{\encodingdefault}{\sfdefault}{bx}{n}

\def\gN{{\mathcal{N}}}

\newcommand{\R}{\mathbb{R}}

\newcommand{\lr}{\alpha}

\def\pa{\partial}

\def\bLam{\bm{\Lambda}}

\def\bu{\mathbf{u}}

\def\baf{\bm{\alpha}}
\def\bsig{\bm{\sigma}}

\def\model{\bm{\phi}_\theta}

\def\scaling{\exp{(-\mLambda \cdot \tau_t)}}
\def\scalingsq{\exp{(-2\mLambda \cdot \tau_t)}}
\def\ueps{\vu_\vepsilon}
\def\ux{\vu_\vx}
\def\uz{\vu_\vz}
\def\PSR{PDE-SpectralRefiner}

\theoremstyle{plain}

\newtheorem{proposition}{Proposition}
\usepackage{indentfirst}
\usepackage{soul}
\usepackage{bm}
\usepackage{extarrows}
\usepackage{threeparttable}
\usepackage[table]{xcolor}

\usepackage{newfloat}
\usepackage{listings}
\DeclareCaptionStyle{ruled}{labelfont=normalfont,labelsep=colon,strut=off} % DO NOT CHANGE THIS
\floatstyle{ruled}
\newfloat{listing}{tb}{lst}{}
\floatname{listing}{Listing}
\title{PDESpectralRefiner: Achieving More Accurate Long Rollouts with Spectral Adjustment}
\author{
    Li Luo
}
\affiliations{
    \textsuperscript{\rm 1}Sun Yat-sen University\\
    511316978@qq.com
}

\usepackage{bibentry}
\begin{document}

\maketitle

\begin{abstract}
% TODO: rewrite
%Recently, \citep{pderefiner} develop a model called PDERefiner that can refine outputs for every step. 
Generating accurate and stable long rollouts is a notorious challenge for time-dependent PDEs (Partial Differential Equations). 
Recently, motivated by the importance of high-frequency accuracy, a refiner model called PDERefiner utilizes diffusion models to refine outputs for every time step, since the denoising process could increase the correctness of modeling high frequency part.
For 1-D Kuramoto-Sivashinsky equation, refiner models can degrade the amplitude of high frequency part better than not doing refinement process. 
However, for some other cases, the spectrum might be more complicated. 
For example, for a harder PDE like Navior-Stokes equation, diffusion models could over-degrade the higher frequency part. 
This motivates us to release the constraint that each frequency weighs the same. 
We enhance our refiner model with doing adjustments on spectral space, which recovers Blurring diffusion models. 
We developed a new v-prediction technique for Blurring diffusion models, recovering the MSE training objective on the first refinement step. 
We show that in this case, for different model backbones, such as U-Net and neural operators, the outputs of PDE-SpectralRefiner are more accurate for both one-step MSE loss and rollout loss. % TODO

\end{abstract}


\section{Introduction}

% \textbf{Notice: Paragraph below not polished.}

% A natural way to analyze PDEs is on reality space. 
% In the past, people often used classical numerical method, such as finite element method (FEM) \citep{FEM} and finite volume method (FVM) \citep{FVM}, \textit{etc}. 
% But these methods require very accurate calculation, and lack of self-adjustable capability. 
Partial differential equations (PDEs) act as an indispensable branch in modern science field. 
As the prosperity of deep learning PDE surrogates, many deep learning models and techniques are developed, and equipped by PDE solvers \citep{zhang2023artificial}. 

One way to learn those PDEs is forward modeling \citep{zhang2023artificial}. 
Given initial conditions or previous observation, models predict quantities in the future. 
A very straightforward approach is that, we let the model take the quantities in the past as input, and the output quantities at next time naturally are used to be the next input of the model for the next time step \citep{towards}. % TODO: polish
But this could be problematic. 
For a system with many timesteps, the error of the model's output could accumulate through time evolution. 
This problem is called stable \textit{rollout}.
Frontiers developed a series of methods to alleviate this issue. \citep{sanchezgonzalez2020learning} inject adversarial Gaussian noise into their model at training step, making the model more robust to the errors for every step.
\citep{brandstetter2022message} naturally regard the output of first bunch of steps as the adversarial perturbed input, then predict the next step behind, mitigating the distributional shift. Temporal bundling, also developed by \citeauthor{brandstetter2022message}, is a nice way to reduce the accumulative error by predicting multiple steps at one time.

% MODIFY BELOW
% There are some methods that can operate on frequency spaces. For classical methods, we can use spectral methods to transform the space of PDE equation into frequency space, separating each frequency modes. 
% It is important for us to research on frequency space since many inherently physical quantities are related to the Fourier space. 
% For example, the amplitude of each frequency modes shows the intensity in specific frequency of this signal. 
% Moreover, the derivative of some quantity with respect to a specific dimension, denoted as $x$, can be obtained accurately through the application of Fourier transform techniques
% \footnote{\url{https://en.wikipedia.org/wiki/Fourier_transform}}. 
% % TODO: link?

% Recently, an increasing number of deep learning methods are focusing on Fourier space, \citet{li2021fourier} develop a model called Fourier neural operator (FNO) that can learn mappings between functions or fields directly in the Fourier space. 
% Since the frequency domain separates high and low frequencies, we can simultaneously learn both large-scale and small-scale information. Mostly, a popular training objective, \textbf{m}ean-\textbf{s}quared \textbf{e}rror (MSE) loss, mostly focuses on low frequency modes, so \citet{pderefiner} proposed a method, which uses diffusion models to \textit{refine} the outputs of the model for every step. 

Modeling a correct spectrum is also a key point for stabling rollout. 
On forward modeling, a widely used loss function is \textbf{m}ean-\textbf{s}quared \textbf{e}rror (MSE) training loss. 
But this type of loss is better on capturing low-frequency errors, which has less focus on high frequencies. 
This could cause an error that could propagate through time evolution.
For 1-D Kuramoto-Sivashinsky (\textbf{KS}) equation, usual MSE training could have a rough output but with a higher amplitude on high frequencies than ground truth \cite{pderefiner}. 
This indicates that MSE training could not have enough capability on modeling high frequencies, corresponding to finer details.
Since diffusion models primarily focus on high-frequency modes,  modulating these modes by reducing their amplitudes \citep{IHDM}, with outputs that have a more correct spectrum, which makes them as a excellent refiner model \citep{pderefiner}. 
On the first step, the refiner model have a rough output. During the refine process, this refiner model gradually outputs closer to the state on the next step. 
\citeauthor{pderefiner} give a very thorough analysis on solving 1-D Kuramoto-Sivashinsky equation, to show that, compared to the outputs from traditional MSE training, the outputs of diffusion models exhibit lower amplitudes in the high-frequency components, making them better aligned with the original data, which have a very low amplitude high frequencies. 

But diffusion models can only control the spectrum by adding noise with different scales. In other words, the denoising process gradually degrades the higher frequency part \textit{implicitly}. 
We experimentally found that for a more complicated problem, for example, Navier-Stokes (NS) equation, has a overall higher frequency spectrum. 
In this case, the output of refiner model could have an overall lower amplitude across whole spectrum.

% DELETED?
For some PDEs which are more complicated, for example, Navior Stokes (NS) equation, the spectra of different physical quantities vary with different spatial dimensions. 
The original diffusion model does not account for these differences, which may increase the difficulty of training. 

This motivates us to control the higher frequency components more directly. % We propose a method that could 
We seek to control the denoising process with different \textit{weights}. 
Specifically, for each refine process, we can operate a Fourier transform, then \textit{multiply each with a coefficient of which each frequency part differs}. 
Since the low frequencies are also mainly focused by refiner model, we don't adjust this part much. 
If we want the noise to affect the high frequencies more, we can multiply a coefficient less than $1$, making the noise weigh more and affect more, vice versa. 
We note that this recover blurring diffusion models \cite{blurring}.
\citet{blurring} explicitly separate each frequency modes by doing forward and backward process in Fourier space, with introducing another schedule to adjust mean coefficient called blurring schedule. 
This technique could help model vectorize the noise schedule, lower the high frequency details of \textit{data}, and accelerate the destruction of higher frequencies more quickly than the original diffusion models, and with particular blurring schedule chosen. 
On one hand, if using the same blurring schedule with \citep{blurring}, we found that the reduction rate of the model's high-frequency amplitude is faster than that of denoising diffusion probabilistic model (DDPM)\citep{ho2020denoising}, which enables quicker reconstruction of high-frequency components. 
On the other hand, if we use a different schedule to make the noise affect less, on the sampling process, we could obtain a quicker degrade. % modify
This provides valuable insights into how to handle spectra. 
% What's more, for more complicated PDE equation like NS equation, we can highlight the differences in the spectra of different physical quantities across various spatial dimensions by manually designing different scaling factors for each quantity. 

V-prediction is a prediction strategy that DDPM predicts velocity moving from data distribution to noise distribution \citep{PGD}. 
By designed noise schedule and v-prediction technique, the first \textit{refinement}/\textit{diffusion} step of diffusion-based refiner model outputs a rough prediction of the next \textit{time} step, resuming MSE training. 
In our case, for different frequency components, the velocity that moves from data to noise is different. 
Additionally, adopting scaling and blurring schedule breaks the variance-preserving property of the original DDPM, which is the key to v-prediction. 
To resume MSE training, we propose a new v-prediction technique that accurately calculates the velocity for each refinement step, aligning with our model setup. 

In summary, we propose PDESpectralRefiner, with following contributions: 
(1) We adopt blurring diffusion models, which can explicitly adjust the amplitude of frequencies. To facilitate the recovery of higher frequencies, we can manually design our blurring schedule, adjusting the varying importance of high frequencies over time. Moreover, our design can take into account the differences in spectra across various physical quantities, which beyond the schedule of \citeauthor{blurring} adopt.
(2) We propose a new technique, \textit{blurring v-prediction technique} that can predict velocity more properly. In conjunction with designed noise schedule, this technique enables us to recover MSE training objective. This technique allows us to more precisely corrupt data to different noise distributions across varying frequencies. 
(3) We show that for several PDE problems, PDESpectralRefiner can recover high frequencies better than DDPM-based PDERefiner. This provides an excellent analytical or application perspective for different PDE problems with different spectra.
\if0
\begin{enumerate}
    \item We adopt blurring diffusion models, which can explicitly adjust the amplitude of frequencies. To facilitate the recovery of higher frequencies, we can manually design our blurring schedule, adjusting the varying importance of high frequencies over time. Moreover, our design can take into account the differences in spectra across various physical quantities.
    \item We propose a new technique that can predict velocity more properly. In conjunction with designed noise schedule, this technique enables us to recover MSE training objective. This technique allows us to more precisely corrupt data to different noise distributions across varying frequencies.
    \item We show that for several PDE problem, PDESpectralRefiner can recover high frequencies better than DDPM-based PDERefiner. This provides an excellent analytical or application perspective for different PDE problems with different spectra.
\end{enumerate}
\fi

\section{Related Work}

\subsection{Stable Long Rollouts}

Long-term stability is one of the most challenging problem on solving PDEs, since the error could accumulate during rolling out, so many PDE researchers have develop strategies to overcome it. Some method focus on mesh-reshaping. \citet{pfaff2021learning} automatically detected the grid shape by adopting remesher for triangular meshes. \citet{wu2023learning} adopted another policy called actor-critic to learn the coarsening of the spatial mesh. Another relatively popular approach is to perturb the data and so forth, increasing the amount of data seen during training, thereby reducing the model's sensitivity to noise perturbations. 
\citet{sanchezgonzalez2020learning} and \citet{pfaff2021learning} added noise to the original data during training to make the output more robust to noise disturbances. \citet{brandstetter2022message} introduced a pushforward trick that only propagates loss backward through a few unrolling steps, alleviating distribution shift problem. \citet{pderefiner} incorporated a diffusion model, formulating the refinement for every step as denoising of diffusion models. In this work we further explore this property in a more complex setting.

\subsection{Diffusion Models}

Diffusion models (DDPM) \citep{ho2020denoising} are powerful generative models that aim to learn the original data distribution through a forward and backward process. Several changes have been made to improve the original DDPM framework. \citet{DDIM} introduced a non-Markovian sampling process that accelerates sampling while maintaining sample quality. \citet{songSDE} unified another model \citep{song2019generative} with DDPM under the framework of stochastic differential equations (SDEs). \citet{IDDPM} extended the original model by predicting the variance of the backward process. However, \citet{analyticdpm} proved that the optimal variance is given by the learned score. 

Originally, diffusion models were widely used in computer vision (\cite{diffusionbeatsgan}, \cite{DALLE}). Recently, many other tasks have adopted diffusion models because they are powerful tools for learning the original data distribution. \citet{hoogeboom2022equivariant}, \citet{bao2023equivariant}, \citet{jing2022torsional}, and \citet{xu2022geodiff} applied diffusion models to generative tasks in the molecular domain. \citet{pderefiner} and \citet{wu2024compositional} used diffusion models to enhance forward modeling and solve inverse problems in PDE tasks.

\section{Preliminaries} % diffusion?

In this section, we provide the preliminary to help readers understand our work easily. In Section \ref{PDEdef}, we give a definition for forward modeling of PDE problems. In Section \ref{diffu}, we briefly introduce diffusion models and blurring diffusion models; In Section \ref{pderef}, we introduce our baseline model PDERefiner, a models that can refine the output of models based on diffusion process.

\subsection{Problem Definition} \label{PDEdef}

\paragraph{Partial Differential Equations.} We focus on time-dependent PDEs in a single time dimension $t \in [0, T]$ and possibly multiple spatial dimensions $\mathbf{x} = [x_1, x_2, ..., x_n] \in \mathbb{X}$ in a bounded domain with $n$ dimension. Solution $\mathbf{u}: [0, T] \times \mathbb{X} \rightarrow \mathbb{R}^n$ and respective derivatives are related through the corresponding PDEs, with $\mathbf{u}(t=0, \mathbf{x})$ are initial conditions at time $t = 0$ and $\mathbb{B}[\mathbf{u}](t,\mathbf{x}) = 0$ are boundary conditions where $\mathbb{B}$ is a boundary operator for $t \in \mathbb{R}^+$ and $\mathbf{x} \in \partial \mathbb{X}$ is on the boundary $\partial \mathbb{X}$ of domain $\mathbb{X}$. Overall, these can be written as:
\begin{align}
    \partial_t \textbf{u} &= F(t, \textbf{x}, \textbf{u}, \partial_\textbf{x} \textbf{u}, \partial_{\textbf{x}\textbf{x}} \textbf{u},...) \,,\\
    \mathbf{u}(t=0, \mathbf{x}) &= \mathbf{u}^0(\mathbf{x}), \qquad \mathbb{B}[\mathbf{u}](t,\mathbf{x}) = 0 \,.
\end{align}
\paragraph{Forward Modeling.} We focus on forward problems for PDE solving \cite{zhang2023artificial}. Given the initial condition or previous observations, the model $\model$ needs to predict the next $T$ solutions based on the first $k$ states. The first $k$ states are called time history $k$, and the next $T$ solutions are called time future. For simplicity, we set both $k$ and $T$ to 1. Forward Modeling can be written as:
\begin{equation}
    \textbf{u}(t + \Delta t) = \model(\textbf{u}(t), t).
\end{equation}
We feed the prediction $\textbf{u}(t + \Delta t)$ to the model $\model$ to obtain predictions at time $t + 2 \Delta t$. This process is called rollout or unrolling the model. The goal for forward modeling is to keep the predictions as close to the original data, as the model unrolling.

\subsection{Diffusion Models and Blurring Diffusion Models}   \label{diffu}

Diffusion models \cite{ho2020denoising, sohl2015deep} are a class of generative models that have gained significant attention for their ability to produce high-quality samples across various domains.
These models work by gradually transforming data from a simple distribution (often Gaussian) into a complex data distribution through a reverse Markov process.

The process starts with data $x_0$ sampled from the data distribution $p(x_0)$. 
The forward process, also known as the diffusion process, gradually adds noise to this data over $T$ timesteps until the data is transformed into a nearly isotropic Gaussian distribution $\mathcal{N}(0, \mathbf{I})$. This can be written as:
%
% \begin{equation}
%     \vz_t = \alpha_t \vx + \sigma_t \vepsilon, \quad \vepsilon \sim \mathcal{N}(0, \mathbf{I})
% \end{equation}
%

% where $\alpha_t$ is decreasing and $\sigma_t$ is a small noise level at time $t$, increasing as $t$ grows up. From a probabilistic perspective, for the forward process, we have:

% $$ , $$
\begin{equation}
    q(\vz_t|\vx) = \gN (\vz_t|\lr_t \vx, \sigma_t^2 \rmI),
\end{equation}

The reverse process aims to reconstruct the original data from the noisy data. 
This involves learning a parameterized model $p(\vz_s|\vz_t)$ that approximates the reverse of the diffusion process for $0 \leq s < t \leq 1$. 
The goal is to learn the distribution of the original data by denoising over several steps. 
This process can be expressed as:
\begin{equation}
    q(\vz_s|\vz_t, \vx) = \gN (\vz_s|\vmu_{t \rightarrow s}(\vz_t, \vx), \sigma_{t \rightarrow s}^2 \rmI),
\end{equation}
where $\sigma_{t \rightarrow s}^2 = (\frac{1}{\sigma_s^2} + \frac{\alpha_{t|s}^2}{\sigma_{t|s}^2})^{-1}$ and $\vmu_{t \rightarrow s}(\vz_t, \vx) = \sigma_{t \rightarrow s}^2(\frac{\alpha_{t|s}^2}{\sigma_{t|s}^2} \vz_t + \frac{\alpha_{s}^2}{\sigma_{s}^2} \vx)$.

The most commonly used loss function is the simplified score matching loss:
\[ L(\theta) = \mathbb{E}_{t, x_0, \epsilon} \left| \epsilon - \epsilon_{\theta}(x_t, t) \right|^2 , \]
where $\epsilon_\theta(x_t, t)$ is a neural network specifically trained to predict the noise added at step $t$.

\cite{PGD} proposed another way to training and sampling diffusion models. 
Rather than predicting $\vx$ or $\vepsilon$, they proposed another strategy that predicting the velocity or movement $\vv$ from $\vx$ to $\vepsilon$, named \textit{v-prediction}. 
As $\alpha_t^2 + \sigma_t^2 = 1$, we can regard $\vx$ and $\vepsilon$ lying on the same ``circle'', with \textit{constant} ``radius'' $\vone$. 
As for a circular motion, $\vv$ is perpendicular to $\vx$, we can easily derive $\vv_t = \alpha_t \vepsilon - \sigma_t \vz_t$. 
On sampling, we can also derive predicted sample $\vx$ by current sample $\vz_t$ and $\vv$: $\vx = \alpha_t \vz_t - \sigma_t \vv$. 
Note that they derive $\vv$ by only taking derivative \textit{w.r.t} $\phi = \arctan{(\sigma_t/\alpha_t)}$. 
Also note that for the ``radius'' is not constant, so we need some modification to have a better training and sampling quality. 
We will discuss this in Section~\ref{PSR}.

\subsection{Blurring diffusion models} 
Diffusion models have been extended and modified in various ways, including conditional generation, improved sampling efficiency with techniques like denoising diffusion implicit models (DDIM)~\citep{DDIM}, and applications in inverse problems \citep{wu2024compositional}. 
Note that in this work, we can regard PDESpectralRefiner as conditional generation, condition on history observations, to generate future states. 

\citep{IHDM} proposed a new type of generative model. Rather than diffusion process, they learn data distribution through heat dissipation. 
\citep{blurring} proved that this type of generative model is also diffusion model, but adding and subtracting noise in Fourier space. 
They choose a schedule that could have a quicker reduction rate of the model's high-frequency part.
\citet{blurring} multiply a vector $\exp{(- \bLam \tau_t)}$ to $\alpha_t$, where $\tau_t = \frac{\sigma_{B}^2}{2} \sin^4{(\frac{\pi t}{2})}$ is called blurring schedule and $\bLam$ is called \textit{scaling}. 
And $\sigma_{B}$ is the maximum blur, which is a tune-able hyperparameter.
The scaling $\bLam$ is a vector that could scale different quantities of different frequencies with different speed. 
\citeauthor{blurring} designed scaling $\bLam$ to be the sum of the frequencies squared for each point. 
This setting has the high frequencies of $\vx_t$ decaying, then high frequencies of noise become larger relatively. 
Therefore, high frequencies of noise has a higher weight, and destructs high frequency of data more quickly. 
But for some other cases, for instance, as we show on solving the 2D Navior-Stokes task, this could over-reduce the high frequency part.
This motivates us designing a new blurring schedule that could alleviate this issue.
In other words, \textit{the coefficients of each frequency modes could be adjusted manually}, according to the specific PDE task we need to solve. 
Depending on different datasets, we can roughly determine various coefficients.

The forward and backward process of blurring diffusion models can be written as:
\begin{align}
    q(\vu_t|\vu_s) &= \gN (\vu_t|\baf_t \vu_s, \bsig^2 \rmI),  \label{blurforward}\\
    q(\vu_s|\vu_t, \vx) &= \gN (\vu_s|\vmu_{t \rightarrow s}, \bsig_{t \rightarrow s}^2 \rmI),    % (\vu_t, \vu_s)
\end{align}

with $\vd_t = (1 - d_{\text{min}}) \cdot \exp{(- \bLam \tau_t)} + d_{\text{min}}, \baf_t = \lr_t \cdot \vd_t, \bsig_t = \sigma_t \cdot \vone$. Note that $\vu_s = \rmV^T \vx$ is the Fourier transform of data $\vx$. Mean and variance in the backward process is almost the same; see the followings: 
\begin{align}
    \bsig_{t \rightarrow s}^2 &= (\frac{1}{\bsig_s^2} + \frac{\baf_{t|s}^2}{\bsig_{t|s}^2})^{-1}, \text{ and } \\
    \vmu_{t \rightarrow s} &= \bsig_{t \rightarrow s}^2(\frac{\baf_{t|s}^2}{\bsig_{t|s}^2} \vu_t + \frac{\baf_{s}^2}{\bsig_{s}^2} \vu_x).   % (\vu_t, \vx)
\end{align}
After vectorizing $\baf$'s and $\bsig$'s, we can control the weights of those frequencies in parallel.

\subsection{PDERefiner} \label{pderef}

One simple and straightforward training for forward PDE modeling is MSE training. 
Recently, PDERefiner utilized diffusion models that can generate the next timestep output by conditioning on previous output or outputs. 
Through designed $\alpha_t$'s and v-prediction \citep[pg.5]{pderefiner}, they resume MSE training objective as the first step of model is a rough output of the next timestep. 
Then the later step of diffusion models are refinement steps. 
As the refinement step increases, model gradually predicts the denoising output, and more and more close to data distribution.
Keeping the accuracy of low frequencies, in terms of fitting accuracy at high frequencies, this method is superior to MSE training. 
% The relationship of DDPM and PDERefiner is that: in forward process, DDPM diffuses from data, representing diffusion time 0, to noise, representing diffusion time 1. 
% In backward process, DDPM sampled a noise from Gaussian distribution at diffusion time 1, gradually denoising as diffusion time decreases to 0. 
% PDERefiner outputs a coarse result of the next \textit{real time} step, at the first \textit{refinement} step 0. 

% \section{Modeling Frequency Spectrum}
% 
\begin{figure}
    \centering
    \includegraphics[width=0.98\linewidth]{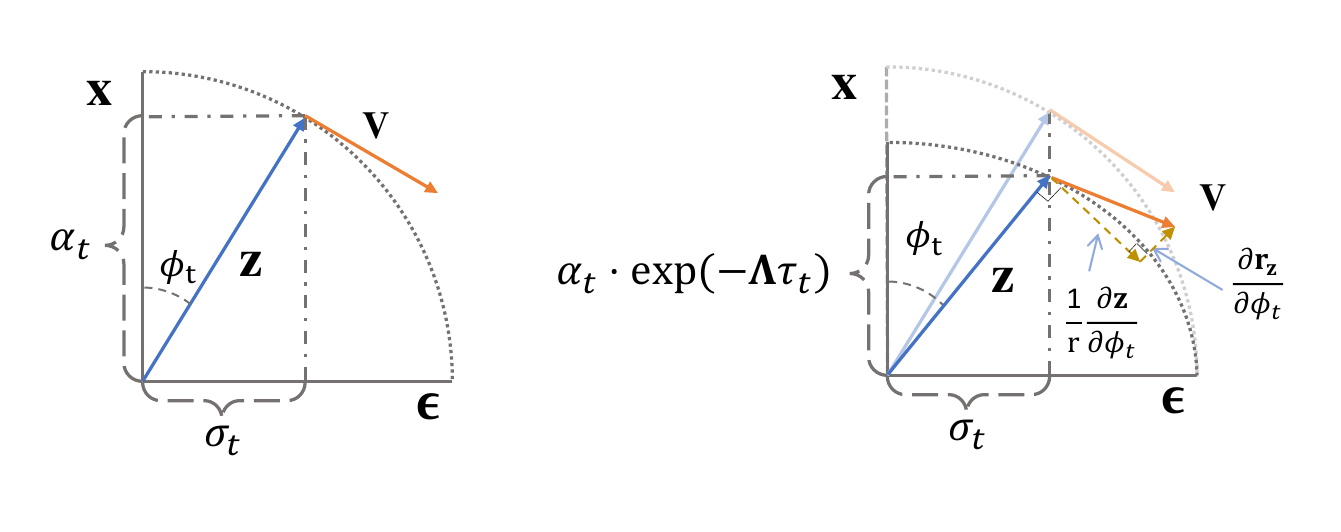}
    % \vspace{-3em}
    \caption{Visualization of reparameterizing the diffusion process (left) and blurring diffusion process in terms of $\phi_t$ and $\vv_\phi$.}
    \label{fig:vpred}
    % \vspace{-10pt}
\end{figure}

\section{PDESpectralRefiner} \label{PSR}

In this section, we introduce PDESpectralRefiner. 
In order to specify the scaling throughout the whole spectrum, we have:
\begin{align} 
    \vd_t = \exp{(-\bLam \cdot \tau_t)}, \label{taut}\\
    \baf_t = \alpha_t \cdot \vd_t, \qquad \bsig_t = \sigma_t \cdot \vone, \label{alphasigma}
\end{align}
Note that this is the same as the blurring diffusion models, which attaches scaling to $\alpha_t$'s, and choosing $\tau_t$ to be equal to $\frac{\sigma_{B}^2}{2} \sin^4{(\frac{\pi t}{2})}$. 
The effect of this setting increments the higher frequencies of $\rvu_\rvx$, and lowers the weights of high frequencies of the noise, therefore corrupts high frequency part of $\vu_\vx$ quickly. 
For further illustration, from the forward process \ref{blurforward}, we have: $\vu_t = \baf_t \vu_s + \bsig \vu = \vd_t \cdot \alpha_t \cdot \vu_s + \bsig \vu$ with $\vu \in \gN (0, \vone)$. 
We can see that, for the scaling $\vd_t \leq 1$ from equation \ref{taut}, the proportion of the noise is larger than DDPM case on the high frequency part, which indicates that high frequencies are destructed more quickly.
But for some cases, on the sampling process, high frequencies might be over-degraded.
Motivated by this, we utilize another schedule, with $\vd_t = \exp{(\bLam \cdot \tau_t)}$. 

We can illustrate it as controlling the weights of those frequencies within a single refinement step. 
For the scaling that could scale the higher frequency up, the noise would have a shorter portion, slowing the diffusion process down. 
For the scaling that could scale the higher frequency down, the noise would have a larger portion, accelerating the diffusion process.

\paragraph{Generality.}
If the scaling is $\vone$ for every frequencies and for all diffusion step, we can recover the DDPM case.
By introducing a blurring schedule the performance of refiner model could be further improved with introducing and detailed tuning.

% We further explore the case which attaches scaling to $\sigma's$. In this case, the noise varies across each dimension, making itself \textit{non-isotropic}. 
% We will explore both case, for scaling high frequencies of the noise up or scaling them down, with an adjustable parameter $\vd_t$.  %TODO: Modify this
% We can easily extend the previous proposition we have derived. 
% More details are at Appendix.

\subsection{Blurring v-prediction}

% To resume MSE training objective, PDERefiner adopts v-prediction technique~\citep{PGD}. 
V-prediction technique~\citep{PGD} is very necessary for refiner models in order to resume MSE training objective.
To be more precise, refiner model should output a rough prediction at the first step.
This could be achieved by v-prediction technique for diffusion models \cite{pderefiner}.
We can see that more clearly through the form of velocity: $\vv_t = \alpha_t \vepsilon - \sigma_t \bu (t) = \sqrt{1 - \sigma_t^2} \vepsilon - \sigma_t \bu (t)$. 
As the $\sigma_0 = 1$, the first step prediction is indeed simple MSE training objective. 

The key point for the establishment of this formula is that $\alpha_t^2 + \sigma_t^2 = 1$. \citet{PGD} depict the relationship among $\vv$, $\vepsilon$ and $\bu$ vividly through a circular motion. 
But for Blurring diffusion model, this property no longer holds as $\baf_t^2 + \bsig_t^2 = \alpha_t^2 * \scalingsq + \sigma_t^2 * \vone ^2 \neq 1$.
Following the same intuition of \citep{PGD}, we can regard the \textit{trajectory} of the transitions from sample distribution to noise distribution as \textit{elliptic}, as the figure \ref{fig:vpred} illustrates.
For simple DDPM, the trajectory is circle, with constant radius $1$, whereas for Blurring diffusion model, $\vr_t$ is equal to$\sqrt{\alpha_t^2 \exp{(-2\mLambda \cdot \tau_t)} + \bsig_t^2}$ varying through diffusion step (see figure \ref{fig:vpred}). 
To alleviate this issue, we propose a new \textit{blurring v-prediction} technique, which recovers simple v-prediction for DDPM:
% So in this case, the v-prediction scheme should be recalculated:
% 
% \begin{equation} \label{newv}
%     \vu_\vz = \baf_t \vu_\vx + \bsig_t \vu_\vepsilon \,.
% \end{equation}
%

% 
\begin{proposition} \label{prop1}
    For PDESpectralRefiner, as $\vr$ changes over time, we have updated $\vv$:
    \begin{align} \label{velo}
        \vv_t = \frac{1}{\vr} \cdot (\alpha_t \ueps - \sigma_t \cdot \scaling \ux) + \\
        \frac{1}{2 \vr^2}\frac{\partial \vr_t^2}{\partial \phi_t}(\alpha_t \cdot \scaling \ux + \sigma_t \ueps),
    \end{align}
    where 
    \begin{equation}
        \begin{aligned}
            \frac{\partial \vr_t^2}{\partial \phi_t} &= 2\alpha_t\sigma_t [1 - \scalingsq + \frac{\alpha_t^2}{\sigma_t^2} \\ 
            & \sigma_B^2 \cos^3{(\frac{t\pi}{2})} \sin{(\frac{t\pi}{2})} \frac{\pi}{\ln{\sigma_0}} \mLambda \scalingsq ]
        \end{aligned}
    \end{equation}
\end{proposition}

Derivations are given at Appendix. 
Note that we could easily extend this result to blurring schedule with $\baf_t = \alpha_t \cdot \sin{(\frac{t\pi}{2})}$, which \cite{blurring} adopted. % \citep{PGD} given
The first term is almost the same as v-prediction for DDPM: $\frac{\pa \uz}{\pa \phi_t} = \alpha_t \ueps - \sigma_t \cdot \scaling \ux = \baf_t \ueps - \bsig_t \ux$. 
This part of velocity is pointing at the perpendicular direction with $\uz$. 
But as the radius $\vr_t$ varying, we need to consider the velocity at the direction of $\uz$. 
The derivative of radius \textit{w.r.t.} $\phi_t$: $\frac{\pa \vr_{t,\perp}}{\pa \phi_t}$ is equal to $\ve_{\vr_t} \frac{\pa \vr_t}{\pa \phi_t} = \frac{1}{2 \vr} \frac{\pa \vr_t^2}{\pa \phi_t} \ve_{\vr_t}$, with $\ve_{\vr_t}$ equal to the unit vector pointing to the same direction as $\uz$. 
The velocity will be pointing all the way along the smooth transition as $\frac{\pa \vr_{t,\perp}}{\pa \phi_t}$ included. 
Travelling along these directions could make it diffuse towards target distribution.

% Although these coefficients look complicated, we can compute it in advance and reduce the training time.

\paragraph{Resuming MSE training objective.}

% 
% $\vr_1 = \sqrt{\vzero^2 \exp{(-2\mLambda \cdot 0)} + \vone^2} = \vone$
% ... = (1 - \scaling + \frac{\alpha_t^2}{\sigma_t^2} \sigma_B \cos^3{(\frac{t\pi}{2})} \sin{(\frac{t\pi}{2})} \frac{1}{\ln{\sigma_0}} \mLambda \scaling)
We can verify that, although the Equation \ref{velo} is quite intricate, this formula can also recover MSE training, with particular blurring schedule $\tau_t = \frac{\sigma_{B}^2}{2} \cos^2{(\frac{\pi t}{2})}$, $\vr_1 = \vone$ and $\frac{\partial \vr_t^2}{\partial \phi_t} = -2\alpha_t \times \vzero \times ... = \vzero$, we have $\vv_1 = \frac{1}{\vone} \cdot (0 \cdot \scaling \ueps - \vone \ux) + \vzero = - \ux$.
This means that, the first step of the PDE-SpectralRefiner resumes common MSE prediction objective. 
Notice that the subscript 1 in $\vv_1$ aligns with the time representation in \citep{ho2020denoising}, where the forward process corresponds to time $0$ to time $1$, and vice versa. 

We will also investigate with the blurring schedule $\tau_t = \frac{\sigma_{B}^2}{2} \sin^2{(\frac{\pi t}{2})}$ at Appendix.
At this time $\vv_1 = - \scaling \cdot \ux$.
Although this doesn't recover MSE training objective at the initial prediction, we can regard it as predicting the low frequencies first, and then recover the high frequencies gradually.

\subsubsection{Sampling.} 
In the sampling procedure, we need to recover $\ux$ as the model outputs velocity rather than $\ux$ itself. 
The following proposition gives the formula of $\vu_s$, given $\uz$ and $\vv$:

\begin{proposition} \label{prop2}
    In sampling process of PDESpectralRefiner, we can derive current sample $\ux$ by velocity:
    \begin{equation}    \label{sample}
        \ux = (\alpha_t \uz + \frac{1}{2 \vr_t^2} \frac{\partial \vr_t^2}{\partial \phi_t} \sigma_t \uz - \vr_t \sigma_t \hat{\vv}_t) / \scaling.
    \end{equation}
    Here $\hat{\vv}_t$ is the prediction or output velocity of our model.
\end{proposition}

We have a clear derivation at Appendix. 
Combining with this ingredient, we can regard our model as a refiner model.

Notice that in the sampling procedure of the blurring diffusion \cite{blurring}, they sample Gaussian noise on the spectral space, and directly add it to $\hat{\vmu_{t \rightarrow s}}$, which lies on the spectral space. 
But the noise sampled from real space can not be added directly to the spectral space. 
Instead, we try two different strategies to modify this. 
First, we sample simple Gaussian noise $\vepsilon$, then apply Fourier transform to it. 
Or, we should sample noise from complex Gaussian distribution: $\tilde{\vepsilon}_k \in \mathcal{CN}(0, 1)$. 
The real and image part of $\tilde{\vepsilon}_k$ both obey $\mathcal{N}(0, \frac{1}{2})$, except for the $[0, 0]$ case. 
We can prove that the Fourier transform of a noise $\vepsilon$ obeys complex Gaussian distribution $\mathcal{CN}(0, 1)$. 
We give a proof at Appendix. 
At sampling procedure, we can sample noise directly at spectral space, reducing time cost by reducing the times for calling Fourier transform. 

\paragraph{Fourier analysis.} \label{fourier}

To further illustrate our motivation and analyze the behaviors and differences of \PSR with different settings, we conduct a Fourier analysis here.
The forward process of diffusion models gradually lifts the power spectrum up, with higher frequencies more sensitive than lower ones. 
So for the backward process, diffusion models try to lower the higher frequencies \cite{IHDM}. 
As the power spectrum of Gaussian noise is roughly constant, the signal-to-noise ratios for each frequency part differ from each other.
In the higher frequency regions, the ratio for each frequency is lower since the amplitude is lower.
This means that high-frequency parts are destroyed quicker than low-frequency parts.
Depending on the blurring schedule we choose, blurring diffusion models could have a more concentrate on different parts.
If we choose scaling $\vd_t = \exp{- \Lambda \cdot \tau_t}$, blurring diffusion models would concentrate on the low frequencies more than refiner model using DDPM.
And if we choose scaling $\vd_t = \exp{\Lambda \cdot \tau_t}$, blurring diffusion models would concentrate on the high frequencies more than refiner model using DDPM.
Low frequencies often represent more on global information and high frequencies represent more on local information.

This could induce a trade-off, for different PDE problem we might adopt different strategy.
For some PDE problems, the output of the refiner model could not have a lower spectrum on higher-frequencies.
But for some other problems, the spectrum of the output might be lower than ground truth. 
we can adjust our blurring schedule to fit these issues.
We will analyze this property on the experiment section and Appendix.

% Blurring schedule could cause the destruction of high-frequency information quicker than simple DDPM case.
% Given the power spectrum of a Gaussian noise is constant, when we are adding noise to our sample, the signal-to-noise ratio is lower in high-frequency parts, as we multiply $\vd_t = \exp{-\Lambda \dot \tau_t}$ which is less than $1$ for all frequency components.
% For blurring diffusion models, Gaussian noise could destroy more quickly than DDPM.
% Interestingly, the reason causes reconstruction quicker could be explained through the reversibility of `motion'. 

% TODO: check this line \ref{fig:noise}
% Although, from the Appendix, we can show that directly sampling from spectral space obtains a smoother spectrum than sampling from real space and then apply a Fourier transform to it. 
% The reason the latter is not smooth might be that we are using discrete Fourier transform instead of the continuous Fourier transform (\citet{george1985mathematical}), transforming Gaussian to another Gaussian. 
% Ideally, if we apply continuous Fourier transform to the noise, the spectrum of this noise should be continuous. 
% But as we use discrete Fourier transform to approximate continuous Fourier transform, there might be some problem with discrete error \textit{e.g.}, inadequate frequency resolution.

\section{Experiments}   \label{exper}

We demonstrate the superiority of PDESpectralRefiner on a common set of PDE benchmarks. 
We mainly focus on 2D Navior-Stokes equation. 
Results for solving 1D Kuramoto-Sivashinsky equation are at Appendix.
We will analyze spatial frequency spectra and performance of both PDERefiner and PDESpectralRefiner, to demonstrate that our model could match the spectra better.

\subsection{2D Navior-Stokes equation}

\paragraph{Simulated data.}

We focus on  \textbf{incompressible} Navier-Stokes equations, a very wildly used PDE benchmark. The vector form of PDE is defined as:

\begin{equation}
    \frac{\pa \rv}{\pa t} + \rv \cdot \nabla \rv = \mu \nabla^2 \rv - \nabla p + \rf, \qquad \nabla \cdot \rv = 0,
\end{equation}

\begin{table*}[t]
\centering
\resizebox{0.9\textwidth}{!}{
\begin{threeparttable}
\scriptsize
\begin{tabular}{l c | c c c c}
    \toprule
    \multicolumn{2}{c|}{\shortstack[c|]{\textbf{Navier-Stokes}}}  & \multicolumn{4}{c}{\shortstack[c]{\textbf{Loss}}} \\
    \midrule
    Model & Params(MB) & unrolled loss & one-step MSE loss & scalar loss & vector loss \\
    \midrule
    \rowcolor[gray]{0.8} \multicolumn{6}{c}{\textit{U-Net}} \\
    MSE training                  & 586     & 2.48e-1$\pm$2.7e-3 & 3.55e-3$\pm$1.3e-4  & 1.98e-3$\pm$3.4e-5 &1.56e-3$\pm$3.0e-5\\
    \midrule
    PDE-Refiner - Step 1          & 586     & \underline{2.25e-1$\pm$2.7e-3} & \underline{1.95e-3$\pm$4.1e-5} & \underline{1.01e-3$\pm$2.2e-5} & \underline{9.38e-4$\pm$2.5e-5} \\
    PDE-SpectralRefiner - Step 1 (up)  & 586     & \textbf{2.23e-1$\pm$2.5e-3} & \textbf{1.86e-3$\pm$4.3e-5} & \textbf{9.73e-4$\pm$2.1e-5} & \textbf{8.86e-4$\pm$2.4e-5}  \\
    PDE-SpectralRefiner - Step 1 (down)  & 586     & {2.31e-1$\pm$2.6e-3} & {2.01e-3$\pm$4.4e-5} & {1.03e-3$\pm$2.2e-5} & {9.73e-4$\pm$2.6e-5}  \\
    \midrule
    PDE-Refiner - Step 3          & 586     & 2.45e-1$\pm$3.0e-3 & 2.02e-3$\pm$4.6e-5 & 1.01e-3$\pm$2.2e-5 & 1.01e-3$\pm$2.4e-5 \\
    PDE-SpectralRefiner - Step 3 (down)  & 586     & \textbf{2.34e-1$\pm$2.9e-3} & \textbf{1.86e-3$\pm$3.9e-5} & \textbf{9.41e-4$\pm$2.2e-5} & \textbf{9.28e-4$\pm$2.1e-5}  \\
    \midrule
    % \rowcolor[gray]{0.8} \multicolumn{7}{c}{\textit{FNO with no mode cutoffs}} \\
    % FNO                         & \checkmark    & 13.00    & 6  & 2.906  & 0.4802   & 92.8 \\
    % FNO (PDERefiner)            & \checkmark    & 13.00    & 3.984  & 1.64  & 0.6879  & \textbf{99.3} \\ 
    % FNO (PDESpectralRefiner)    &               & 13.00    & 3.935  & 1.000  & 0.6919 & 85.0 \\
    % \midrule
    \rowcolor[gray]{0.8} \multicolumn{6}{c}{\textit{FNO}} \\
    MSE training                  & 559     & \textbf{3.11e-1$\pm$2.7e-3} & \textbf{7.16e-3$\pm$1.4e-4}  & \textbf{3.02e-3$\pm$6.0e-5} & \textbf{4.15e-3$\pm$7.1e-5}\\
    PDE-Refiner - Step 3          & 559     & 3.93e-1$\pm$3.2e-3 & 1.04e-2$\pm$2.0e-4 & 3.90e-3$\pm$5.6e-5 & 6.44e-3$\pm$1.2e-5\\
    PDE-SpectralRefiner - Step 3  & 559     & \underline{3.88e-1$\pm$3.1e-3} & \underline{1.01e-2$\pm$1.2e-4} & \underline{3.84e-3$\pm$5.6e-5} & \underline{6.31e-3$\pm$1.2e-5}\\
    \bottomrule
\end{tabular}
\end{threeparttable}}
\caption{Results of mean MSE loss on 2D Navior Stokes equation. The longer the time is, more stable the rollout is. For MSE loss, we take the sum across the batch dimension. Bold values show better result on the same model. The best results are highlighted in bold, and the second-best results are underlined. (Up) means we choose scaling $\vd_t = \exp{\Lambda \cdot \tau_t}$ while (down) means choosing scaling $\vd_t = \exp{- \Lambda \cdot \tau_t}$.}
\label{tabletestresults}
% \vspace{-1em}
\end{table*}

\begin{figure}[H]
	\centering
    \begin{minipage}[b]{0.35\textwidth}
        \begin{center}
           % \vspace*{-1em}
           \includegraphics[width=\textwidth]{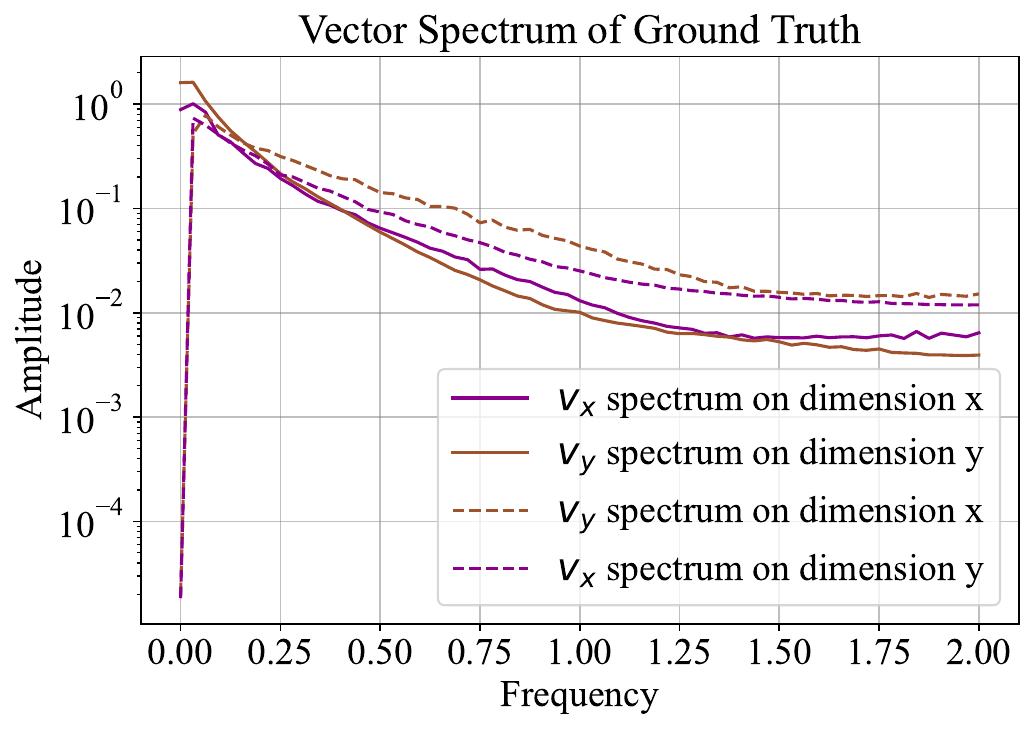}
        \end{center}
        % \vspace{-1.5em}
    \end{minipage}
    \caption{Frequency Spectrum of original Navior Stokes equation. Plots with colors.}
    \label{fig:NS}
    \hspace{5pt}
    \begin{minipage}[b]{0.35\textwidth}
        \begin{center}
        \includegraphics[width=\textwidth]{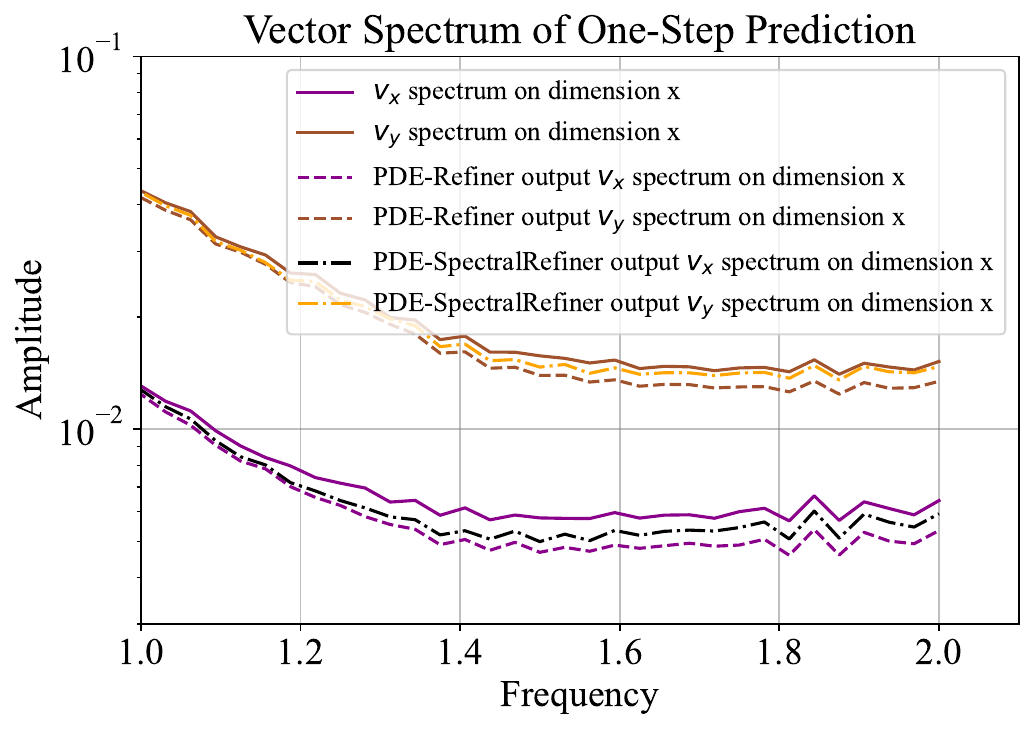}
        \end{center}
        % \vspace{-1.5em}
    \end{minipage}
    \caption{Frequency Spectrum of Navier-Stokes equation prediction for PDE-Refiner and PDE-SpectralRefiner. Both are $1$ refinement step.}
    \label{fig:comparison}
    \hspace{5pt}
    % \begin{minipage}[b]{0.35\textwidth}
    %     \begin{center}
    %         \includegraphics[width=\textwidth]{Styles/figures/comparison_vector_Fp.pdf}
    %     \end{center}
    %     % \vspace{-1.5em}
    %     \caption{Frequency Spectrum of original Navior Stokes equation and PDERefiner for \textbf{U-Fnet} based prediction.}
    %     \label{fig:vector_pure}
    % \end{minipage}
    % \hspace{5pt}
    % \begin{minipage}[b]{0.35\textwidth}
    %     \begin{center}
    %         \includegraphics[width=\textwidth]{Styles/figures/comparison_vector_BlurFunet1.pdf}
    %     \end{center}
    %     % \vspace{-1.5em}
    %     \caption{Frequency Spectrum of original Navior Stokes equation and PDESpectralRefiner for \textbf{U-Fnet} prediction.}
    %     \label{fig:vector_blur}
    % \end{minipage}
    %\vspace{-1em}
    % \caption{Spectral analysis for U-Fnet based prediction}
    % \vspace{-2mm}
\end{figure}

where $\rv$ is the velocity flow fields $\rv: \mathcal{X} \rightarrow \R^2$, $\nabla p$ is the internal pressure and $\rf$ is the external force. 
$\nabla \cdot \rv = 0$ is derived from $\frac{\pa \rho}{\pa t} + \nabla \cdot (\rho \rv) = 0$ where incompressible indicates that $\rho$ is constant, yielding mass conservation. 
This 2D Navier-Stokes data is obtained on a grid with spatial resolution of $128 \times 128$ with $\Delta x$ and $\Delta y$ equal to $0.25$. 
All trajectories have $14$ time points, with $\Delta t = 1.5s$ and total time $21s$. 
We use 2D Navior Stokes dataset with open source \verb+PDEArena+ \citep{towards}, which is generated via \verb+Phi-Flow+ \citep{phiflow}.
% \footnote{\url{https://github.com/pdearena/pdearena}}

% Fourier neural operators(FNO) \citep{li2021fourier}
\subsubsection{Experimental Setup.} 
We adopt modern U-net \citep{towards} as our backbone model. 
We use modern U-net with hidden size $128$ and $3$ downsampling layers. 
Furthermore, we also use Fourier neural operators (FNOs) \citep{li2021fourier} and U-shaped neural operators (UNOs) \citep{rahman2023uno}. 
We will have a clear explanation for modes cutoff at Appendix, also for more results of U-Net, FNO and UNO architectures.
We let the minimum noise standard deviation be $1 \times 10^{-3}$. 
We compare PDESpectralRefiner with PDE-Refiner and MSE training objective with these model architecture. 
We investigate 1 and 3 refinement step for both PDE-Refiner and \PSR.
In Table.\ref{tabletestresults}, we report MSE on the test set. 
We also analyze part of the FNO results here.
In short, FNOs here have hidden size $128$ and $16$ modes for each of the $4$ layers.
% Model architecture and Dataset setting
Since the trajectory length is only $14$, we only use one history as input, and predict one time future. As for the neural network architecture, we control the parameters of these networks to let them have similar model parameters size. 
We have a plot of correlation time and MSE loss for each time-step of the test set. We provide our experimental details at Appendix.

\paragraph{Spectrum for Navior Stokes equation.}
The spectrum for Navior Stokes equation is much more complicated than KS equation. 
As we show in Figure~\ref{fig:NS}, we could see that for Navior Stokes equation, the spectrum for different quantities is different. 
We randomly choose 4 samples, applying FFT to the batch of them, taking the mean value along the whole time and 4 samples, and draw the spectrum of $v_x$ and $v_y$. 
We also take the mean value along the dimension we don't analyze. 
For example, for the spectrum of $v_x$ in the x-direction, we take the mean value along the dimension y. 
In general, we can see that the spectrum of $v_x$ in the x-direction and the spectrum of $v_y$ in the y-direction are lower than the spectrum of $v_x$ in the y-direction and the spectrum of $v_y$ in the x-direction. 
This could cause training more difficult.

\paragraph{Note on the spectrum of 1D KS equation and 2D NS equation. }
2D Navior Stokes equation is a much more difficult problem than 1D KS equation. 
We first introduce the concept of power spectral density (PSD). 
The power spectral density of a signal is a measure that describes how the power of a signal or time series is distributed with frequency. 
This is highly correlated to the amplitude across frequency spectrum. 
For Navior Stokes equation, there exists a phenomenon called energy cascade, this says that the energy in the lower frequencies could propagate to higher frequencies, causing the higher frequencies having more energies. 
The same as KS equation, but the PSD of KS equation decays faster than the NS equation, from low frequency to high frequency \citep{baez2022kuramoto, wittenberg1999scaleKS, temam2001navier}.  % \footnote{\url{https://en.wikipedia.org/wiki/Energy_cascade}}

\paragraph{Analysis.}
We also try different maximum blurring for both scaling $\vd_t = \exp{(- \Lambda \cdot \tau_t)}$ and $\vd_t = \exp{(\Lambda \cdot \tau_t)}$.
For $\vd_t = \exp{(- \Lambda \cdot \tau_t)}$, we set maximum blurring for $[2, 4, 8]$.
For the other case, we set $[2, 4]$.
From table \ref{tabletestresults}, for the U-Net backbone, PDE-SpectralRefiner has the best performance for different maximum blurring hyper-parameter.
For refiner model that only have $1$ refinement step, PDE-SpectralRefiner that has $\vd_t = \exp{(\Lambda \cdot \tau_t)}$ and maximum blurring $8$ could have a better result, whereas for refiner model that have $3$ refinement step, PDE-SpectralRefiner that has $\vd_t = \exp{(-\Lambda \cdot \tau_t)}$  and maximum blurring $4$ performs better.
We will analyze this one by one.
From \ref{fourier}, when the $\baf_t$ are chosen to be scaled up, the model concentrates on high-frequencies more than refiner model based on DDPM.
Since the model only has two output steps, with the initial prediction and only one refinement step, we need to determine the higher-frequency information quicker.
This setting could alleviate this issue, with higher signal-to-noise ratio on high-frequency areas, pushing the model focus on higher frequencies more.
When the $\baf_t$ are chosen to be scaled down, the refiner model has more focus on lower-frequency components.
But on one hand, with more refinement steps, the model could gradually focus on high-frequencies.
On the other hand, from fig \ref{fig:comparison}, the model could over-degrade the high-frequency information.
For refiner model that based on DDPM, the output of the last refinement step has a lower amplitude of higher frequency than the ground truth.
This could be one of the resources of the error.
Meanwhile, from figure, for blurring schedule $\vd_t = \exp{(- \Lambda \cdot \tau_t)}$ and max blur $\sigma_0 = 2$, \PSR could match high frequencies better, because the designed blurring schedule alleviate this issue by not degrading the spectrum of this sample too much. 
Being able to adjust the weight for each frequency components, so the improve the alignment of the spectrum is the key motivation of our paper.

\paragraph{Results for FNO}

Interestingly, from table, we found that for 2D Navier-Stokes equation, with frequency cut-off, adopting refinement process could deteriorate the performance.
% What's more, with more refinement steps, the performance degrades more.
This result is not consistent with the results obtained in \cite{pderefiner}, whose experiments were conducted on 1D KS equation. 
From the section \ref{fourier} and the experiment section on the main body, we illustrate the difference between these two PDE problems.
In short, the power (amplitude) in high-frequencies for 2D Navier-Stokes equation is higher than 1D KS equation.
Given the limited model size, refiner models based on FNOs could not generate the right power spectrum, which can be reflected on both correlation score in every step.
From fig at Appendix, one probable explanation could be: adopting diffusion process makes FNOs with mode cut-off hard to focus on high-frequency information.
FNOs focus on the interaction among different frequencies within the input. 
As the input are noised, FNOs are hard to learn the interaction between different frequency modes, especially with mode cut-off.
For 1D KS equation, high-frequency amplitudes are lower the 2D NS equation overall.
Although the term $uu_x$ induces interaction between low and high frequency components, adding noise doesn't affect this too much.
But for 2D NS equation, adding noise could be break the interaction between low and high frequency interaction.
This issue becomes severe if we have more modes being cut off.

Nonetheless, PDE-SpectralRefiner still outperforms PDE-Refiner in this case, proving that FNOs could have a better coordination with blurring diffusion models with the specific blurring schedule.
For blurring schedule that scale the $\alpha_t$'s up, PDE-SpectralRefiner that based on FNOs could obtain a better result.
Firstly, as we also lift the signal-to-noise ratio, the model could see the signal more, making the interaction more explicit.
Secondly, we lift the high-frequency part more than low-frequency part, emphasizing high-frequency part.
Note that this only changing the amplitude for complex numbers of all modes but not the angle.
So this approach can be considered as spectral \textbf{adjustment}, reweighting each frequency modes. 
With proper spectral adjustment, the model's performance could be boosted.

% ?
% We can see that this model can not recover the amplitude of high frequency part, especially for PDERefiner, which is lower amplitude of higher frequency than the ground truth.
% This indicates a bottleneck from DDPM as refiner model, which may over-degrading the spectrum.
% Meanwhile, from figure, for blurring schedule $\vd_t = \exp{(\Lambda \cdot \tau_t)}$ and max blur $\sigma_0 = 2$, \PSR could match high frequencies better, because the designed blurring schedule alleviate this issue by not degrading the spectrum of this sample too much. 
% Note that this is the key motivation of our paper.

\section{Conclusion}
In this work, we develop PDESpectralRefiner, a new model that can refine every step of solving PDE problems with spectral reweighting. 
Using this model, we can achieve more stable rollout than the previous work that using simple DDPM to do those refinement. 
By giving a spectral analysis, we can show that the output of our model has more correct higher frequencies to the original data than the output of PDERefiner. 
One limitation is that it still can not have a good cooperation with FNOs, even if we do not have a mode cutoff. 
Another limitation is that, for non-uniform grids, we might need to utilize projector other than FFT, which might introduce more complexity.
% And another limitation is that the number of function evaluation is often one to four times than simply regard the output of model as the input of next step.
% And another limitation is that we only consider the amplitude of the imaginary data after Fourier transform, but don't consider angle correspondence.
In summary, we propose a powerful model that can incorporate spectra design, and provides a valuable prospective to analyze frequency spectrum.

% \appendix

% \section{Acknowledgments}

% None.

\bibliography{aaai25}

\end{document}